%% file: main.tex
\pdfoutput=1
\documentclass[12pt,a4paper]{article}
\usepackage{ifthen} % for conditional statements
\newboolean{pdflatex}
\setboolean{pdflatex}{true} % False for eps figures 

\newboolean{articletitles}
\setboolean{articletitles}{true} % False removes titles in references

\newboolean{uprightparticles}
\setboolean{uprightparticles}{false} %True for upright particle symbols

\def\paperauthors{LHCb collaboration} % Leave as is for PAPER, CONF and FIGURE
\def\paperasciititle{Lambda transverse polarization measurement in pNe collisions at sqrt(s)NN = 68.4 GeV with the LHCb detector} % Set ASCII title here !! MAKE sure it's only ASCII characters !! 
\def\papertitle{Transverse polarization measurement of \Lz hyperons in $p$Ne collisions at \sqsnn = 68.4\gev with the \lhcb detector} % Latex formatted title
\def\paperkeywords{{High Energy Physics}, {LHCb}} % Comma separated list
\def\papercopyright{\the\year\ CERN for the benefit of the LHCb collaboration} % new since 9/Apr/2018
\def\paperlicence{CC BY 4.0 licence}
\def\paperlicenceurl{https://creativecommons.org/licenses/by/4.0/}

\input{preamble}
\usepackage{longtable} % only for template; not usually to be used in PAPERs
\begin{document}

%%%%%%%%%%%%%%%%%%%%%%%%%
%%%%% Title     %%%%%%%%%
%%%%%%%%%%%%%%%%%%%%%%%%%
\renewcommand{\thefootnote}{\fnsymbol{footnote}}
\setcounter{footnote}{1}

% %%%%%%% CHOOSE TITLE PAGE--------
%\onecolumn
%\input{title-LHCb-INT}
%\input{title-LHCb-ANA}
%\input{title-LHCb-CONF}
%\input{title-LHCb-FIGURE}
\input{title-LHCb-PAPER}

%\twocolumn
% %%%%%%%%%%%%% ---------

\renewcommand{\thefootnote}{\arabic{footnote}}
\setcounter{footnote}{0}

%%%%%%%%%%%%%%%%%%%%%%%%%%%%%%%%
%%%%%  Table of Content   %%%%%%
%%%%%%%%%%%%%%%%%%%%%%%%%%%%%%%%
%%%% Uncomment if desired
%\tableofcontents
\cleardoublepage

%%%%%%%%%%%%%%%%%%%%%%%%%
%%%%% Main text %%%%%%%%%
%%%%%%%%%%%%%%%%%%%%%%%%%

\pagestyle{plain} % restore page numbers for the main text
\setcounter{page}{1}
\pagenumbering{arabic}

%% Uncomment during review phase. 
%% Comment before a final submission.

%% This is the main body
%% It is useful to have a single file so comemnts are not missed in overleaf.
\input{body}

% Do not include this in any draft (just for information in the template)
%\input{acknowledgements_intro}
% Comment this in for paper drafts; do not include this in analysis note, conference and figure reports
\clearpage
\newpage
\input{acknowledgements}

%%CDA: APPENDICI%%

%\input{supplementary}
%\input{appendix}

% This should be taken out in the final paper
%\input{supplementary-app}

\addcontentsline{toc}{section}{References}
%\setboolean{inbibliography}{true}
\bibliographystyle{LHCb}

%%CDA: BIBLIOGRAFIA%%
\newpage
\newpage
\newpage

\bibliography{main,standard,LHCb-PAPER,LHCb-CONF,LHCb-DP,LHCb-TDR,mybib}

\newpage
\input{Authorship_LHCb-PAPER-2024-009}

\end{document}

%% file: preamble.tex
\usepackage[top=1in, bottom=1.25in, left=1in, right=1in]{geometry}

\usepackage{microtype}
\usepackage{lineno}  % for line numbering during review
\usepackage{xspace} % To avoid problems with missing or double spaces after
\usepackage{caption} %these three command get the figure and table captions automatically small

\usepackage{graphicx}  % to include figures (can also use other packages)
\usepackage{color}
\usepackage{colortbl}
\graphicspath{{./figs/}} % Make Latex search fig subdir for figures
\usepackage{amsmath} % Adds a large collection of math symbols
\usepackage{amssymb}
\usepackage{amsfonts}
\usepackage{upgreek} % Adds in support for greek letters in roman typeset

\newcommand*\patchAmsMathEnvironmentForLineno[1]{%
\expandafter\let\csname old#1\expandafter\endcsname\csname #1\endcsname
\expandafter\let\csname oldend#1\expandafter\endcsname\csname
end#1\endcsname
 \renewenvironment{#1}%
   {\linenomath\csname old#1\endcsname}%
   {\csname oldend#1\endcsname\endlinenomath}%
}
\newcommand*\patchBothAmsMathEnvironmentsForLineno[1]{%
  \patchAmsMathEnvironmentForLineno{#1}%
  \patchAmsMathEnvironmentForLineno{#1*}%
}
\AtBeginDocument{%
\patchBothAmsMathEnvironmentsForLineno{equation}%
\patchBothAmsMathEnvironmentsForLineno{align}%
\patchBothAmsMathEnvironmentsForLineno{flalign}%
\patchBothAmsMathEnvironmentsForLineno{alignat}%
\patchBothAmsMathEnvironmentsForLineno{gather}%
\patchBothAmsMathEnvironmentsForLineno{multline}%
\patchBothAmsMathEnvironmentsForLineno{eqnarray}%
}

\usepackage{hyperxmp}

\usepackage[pdftex,
            pdfauthor={\paperauthors},
            pdftitle={\paperasciititle},
            pdfkeywords={\paperkeywords},
            pdfcopyright={Copyright (C) \papercopyright},
            pdflicenseurl={\paperlicenceurl}]{hyperref}
\usepackage[colorinlistoftodos,textsize=scriptsize]{todonotes}

\usepackage[bottom,flushmargin,hang,multiple]{footmisc}

\usepackage[all]{hypcap} % Internal hyperlinks to floats.

\input{lhcb-symbols-def} % Add in the predefined LHCb symbols

\usepackage{cite} % Allows for ranges in citations
\usepackage{mciteplus}

%% file: lhcb-symbols-def.tex
\usepackage{xspace} 
\usepackage{upgreek}

\def\lhcb   {\mbox{LHCb}\xspace}

\def\belle  {\mbox{Belle}\xspace}

\def\MagUp {\mbox{\em Mag\kern -0.05em Up}\xspace}

\ifthenelse{\boolean{uprightparticles}}%
{

 \def\PDelta      {\ensuremath{\Delta}\xspace}                 
 \def\PXi         {\ensuremath{\Xi}\xspace}                 
 \def\PLambda     {\ensuremath{\Lambda}\xspace}                 
 \def\PSigma      {\ensuremath{\Sigma}\xspace}                 
 \def\POmega      {\ensuremath{\Omega}\xspace}                 
 \def\PUpsilon    {\ensuremath{\Upsilon}\xspace}
 \let\oldPi\Pi
 \def\PPi         {\ensuremath{\oldPi}\xspace}

 \def\PB      {\ensuremath{\mathrm{B}}\xspace}                 
                  
 \def\PD      {\ensuremath{\mathrm{D}}\xspace}

 \def\PK      {\ensuremath{\mathrm{K}}\xspace}

 \def\Pi      {\ensuremath{\mathrm{i}}\xspace}

 \def\Ps      {\ensuremath{\mathrm{s}}\xspace}

 \def\thebaroffset{0.0em}
}
{

 \mathchardef\PDelta="7101
 \mathchardef\PXi="7104
 \mathchardef\PLambda="7103
 \mathchardef\PSigma="7106
 \mathchardef\POmega="710A
 \mathchardef\PUpsilon="7107
 \mathchardef\PPi="7105
                  
 \def\PB      {\ensuremath{B}\xspace}                 
                  
 \def\PD      {\ensuremath{D}\xspace}

 \def\PK      {\ensuremath{K}\xspace}

 \def\Pi      {\ensuremath{i}\xspace}

 \def\Ps      {\ensuremath{s}\xspace}

 \def\thebaroffset{0.18em}
}
\newcommand{\offsetoverline}[2][\thebaroffset]{\kern #1\overline{\kern -#1 #2}}%

\makeatletter
\ifcase \@ptsize \relax% 10pt
  \newcommand{\miniscule}{\@setfontsize\miniscule{4}{5}}% \tiny: 5/6
\or% 11pt
  \newcommand{\miniscule}{\@setfontsize\miniscule{5}{6}}% \tiny: 6/7
\or% 12pt
  \newcommand{\miniscule}{\@setfontsize\miniscule{5}{6}}% \tiny: 6/7
\fi
\makeatother

\DeclareRobustCommand{\optbar}[1]{\shortstack{{\miniscule (\rule[.5ex]{1.25em}{.18mm})}
  \\ [-.7ex] $#1$}}

\def\squark    {{\ensuremath{\Ps}}\xspace}

\def\KorKbar {\kern \thebaroffset\optbar{\kern -\thebaroffset \PK}{}\xspace}

\def\D       {{\ensuremath{\PD}}\xspace}

\def\DorDbar {\kern \thebaroffset\optbar{\kern -\thebaroffset \PD}\xspace}

\def\Dp      {{\ensuremath{\D^+}}\xspace}
\def\Dm      {{\ensuremath{\D^-}}\xspace}

\def\DpDm    {\ensuremath{\Dp {\kern -0.16em \Dm}}\xspace}

\def\B       {{\ensuremath{\PB}}\xspace}

\def\BorBbar {\kern \thebaroffset\optbar{\kern -\thebaroffset \PB}\xspace}

\def\Bd      {{\ensuremath{\B^0}}\xspace}

\def\BdorBdbar {\kern \thebaroffset\optbar{\kern -\thebaroffset \Bd}\xspace}

\def\Bs      {{\ensuremath{\B^0_\squark}}\xspace}

\def\BsorBsbar {\kern \thebaroffset\optbar{\kern -\thebaroffset \Bs}\xspace}

\def\Y#1S{\ensuremath{\PUpsilon{(#1S)}}\xspace}

\def\Lz          {{\ensuremath{\PLambda}}\xspace}
\def\Lbar        {{\ensuremath{\offsetoverline{\PLambda}}}\xspace}
\def\LorLbar     {\kern \thebaroffset\optbar{\kern -\thebaroffset \PLambda}\xspace}

\def\to                 {\ensuremath{\rightarrow}\xspace}

\def\AT#1     {\ensuremath{A_{\mathrm{T}}^{#1}}\xspace}           % 2

\def\C#1      {\ensuremath{\mathcal{C}_{#1}}\xspace}                       % 9
\def\Cp#1     {\ensuremath{\mathcal{C}_{#1}^{'}}\xspace}                    % 7
\def\Ceff#1   {\ensuremath{\mathcal{C}_{#1}^{\mathrm{(eff)}}}\xspace}        % 9  
\def\Cpeff#1  {\ensuremath{\mathcal{C}_{#1}^{'\mathrm{(eff)}}}\xspace}       % 7
\def\Ope#1    {\ensuremath{\mathcal{O}_{#1}}\xspace}                       % 2
\def\Opep#1   {\ensuremath{\mathcal{O}_{#1}^{'}}\xspace}                    % 7

\newcommand{\aunit}[1]{\ensuremath{\text{\,#1}}}       
\newcommand{\tev}{\aunit{Te\kern -0.1em V}\xspace}
\newcommand{\gev}{\aunit{Ge\kern -0.1em V}\xspace}
\newcommand{\mev}{\aunit{Me\kern -0.1em V}\xspace}
\newcommand{\kev}{\aunit{ke\kern -0.1em V}\xspace}
\newcommand{\ev}{\aunit{e\kern -0.1em V}\xspace}
 
\newcommand{\mevc}{\ensuremath{\aunit{Me\kern -0.1em V\!/}c}\xspace}
\newcommand{\gevc}{\ensuremath{\aunit{Ge\kern -0.1em V\!/}c}\xspace}
\newcommand{\mevcc}{\ensuremath{\aunit{Me\kern -0.1em V\!/}c^2}\xspace}
\newcommand{\gevcc}{\ensuremath{\aunit{Ge\kern -0.1em V\!/}c^2}\xspace}
\def\gsim{{~\raise.15em\hbox{$>$}\kern-.85em
          \lower.35em\hbox{$\sim$}~}\xspace}
\def\lsim{{~\raise.15em\hbox{$<$}\kern-.85em
          \lower.35em\hbox{$\sim$}~}\xspace}

\def\sqsnn {\ensuremath{\protect\sqrt{s_{\scriptscriptstyle\text{NN}}}}\xspace}
\def\pt         {\ensuremath{p_{\mathrm{T}}}\xspace}

\def\evtgen     {\mbox{\textsc{EvtGen}}\xspace}

\def\pythia     {\mbox{\textsc{Pythia}}\xspace}

\def\tell1  {TELL1\xspace}
\def\ukl1   {UKL1\xspace}

\newcommand{\lhcborcid}[1]{\href{https://orcid.org/#1}{\hspace*{0.1em}\raisebox{-0.45ex}{\includegraphics[width=1em]{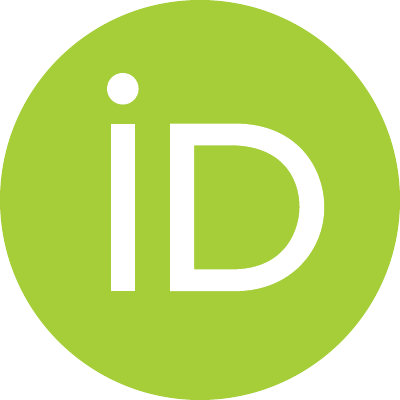}}}}

%% file: title-LHCb-PAPER.tex
% ===============================================================================
% Purpose: LHCb-PAPER journal paper title page template
% Author: 
% Created on: 2010-09-25
% ===============================================================================

%%%%%%%%%%%%%%%%%%%%%%%%%
%%%%%  TITLE PAGE  %%%%%%
%%%%%%%%%%%%%%%%%%%%%%%%%
\begin{titlepage}
\pagenumbering{roman}

% Header ---------------------------------------------------
\vspace*{-1.5cm}
\centerline{\large EUROPEAN ORGANIZATION FOR NUCLEAR RESEARCH (CERN)}
\vspace*{1.5cm}
\noindent
\begin{tabular*}{\linewidth}{lc@{\extracolsep{\fill}}r@{\extracolsep{0pt}}}
\ifthenelse{\boolean{pdflatex}}% Logo format choice
{\vspace*{-1.5cm}\mbox{\!\!\!\includegraphics[width=.14\textwidth]{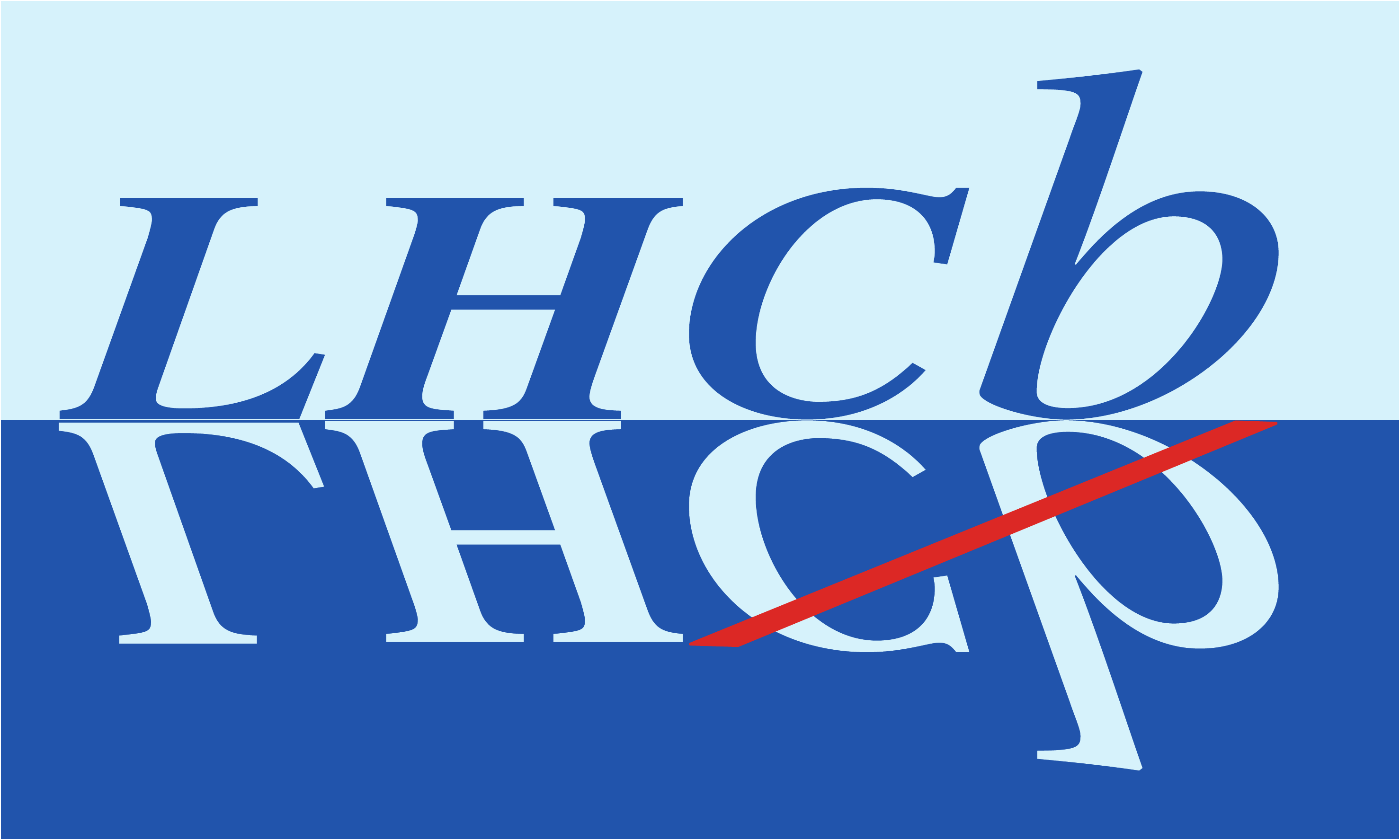}} & &}%
{\vspace*{-1.2cm}\mbox{\!\!\!\includegraphics[width=.12\textwidth]{figs/lhcb-logo.eps}} & &}%
\\
 & & CERN-EP-2024-121 \\  % ID 
 & & LHCb-PAPER-2024-009 \\  % ID 
 & & September 19, 2024 \\ % Date - Can also hardwire e.g.: 23 March 2010
 & & \\
% not in paper \hline
\end{tabular*}

\vspace*{4.0cm}

% Title --------------------------------------------------
{\normalfont\bfseries\boldmath\huge
\begin{center}
% DO NOT EDIT HERE. Instead edit macro in main.tex to keep metadata correct
  \papertitle 
\end{center}
}

\vspace*{2.0cm}

% Authors -------------------------------------------------
\begin{center}
%In the footnote, replace 'paper' by 'Letter' in case of submission to PRL or PLB 
% Edit macro in main.tex to keep metadata correct
\paperauthors\footnote{Authors are listed at the end of this paper.}
\end{center}

\vspace{\fill}

% Abstract -----------------------------------------------
\begin{abstract}
  \noindent
  A measurement of the transverse polarization of the $\ensuremath{\PLambda}$ and $\ensuremath{\offsetoverline{\PLambda}}$ hyperons in $p$Ne fixed-target collisions at $\ensuremath{\protect\sqrt{s_{\scriptscriptstyle\text{NN}}}}$ = 68.4 $\aunit{Ge\kern -0.1em V}$ is presented using data collected by the LHCb detector. The polarization is studied using the decay $\ensuremath{\PLambda}\rightarrow p \pi^-$ together with its charge conjugated process,  
  the integrated values measured are
  \begin{linenomath}$$ P_{\ensuremath{\PLambda}} = 0.029 \pm 0.019 \, (\rm{stat}) \pm 0.012 \, (\rm{syst}) \, , $$ $$ P_{\ensuremath{\offsetoverline{\PLambda}}} = 0.003 \pm 0.023 \, (\rm{stat}) \pm 0.014 \,(\rm{syst}) \,. $$\end{linenomath}
  Furthermore, the results are shown as a function of the Feynman~$x$~variable, transverse momentum, pseudorapidity and rapidity of the hyperons, and are compared with previous measurements.
  
\end{abstract}

\vspace*{2.0cm}

\begin{center}
  Published in JHEP 09 (2024) 082
\end{center}

\vspace{\fill}

{\footnotesize 
% Edit macro in main.tex to keep metadata correct
\centerline{\copyright~\papercopyright. \href{\paperlicenceurl}{\paperlicence}.}}
\vspace*{2mm}

\end{titlepage}

%%%%%%%%%%%%%%%%%%%%%%%%%%%%%%%%
%%%%%  EOD OF TITLE PAGE  %%%%%%
%%%%%%%%%%%%%%%%%%%%%%%%%%%%%%%%

%  empty page follows the title page ----
\newpage
\setcounter{page}{2}
\mbox{~}
%\newpage
%
%% Author List ----------------------------
%%  You need to get a new author list!
%\input{LHCb_authorlist.tex}
%
%The author list for journal publications is provided by the Membership Committee shortly after 'approval to go to paper' has been given.
%%It will be made available on the page
%%\verb!http://www.physik.uzh.ch/~strauman/forMemCo/LHCb-PAPER-XXXX-XXX/! .
%It will be sent to you by email shortly after a paper number has beens assigned.
%The author list should be included already at first circulation, 
%to allow new members of the collaboration to verify whether they have been included correctly.
%Occasionally a misspelled name is corrected or associated institutions become full members.
%In that case, a new author list will be sent to you.
%In case line numbering doesn't work well after including the authorlist, try moving the \verb!\bigskip! after the last author to a separate line.
%
%
%The authorship for Conference Reports should be ``The LHCb
%  collaboration'', with a footnote giving the name(s) of the contact
%  author(s), but without the full list of collaboration names.

%% file: body.tex
\section{Introduction}
The spontaneous transverse polarization of \Lz hyperons, spin 1/2 baryons with a mass of $1115.683 \pm 0.006$ \mevcc~\cite{PDG2022}, was first observed in 1976 in unpolarized fixed-target collisions of protons with an energy of 300\gev and a beryllium target \cite{Bunce:1976yb}. This result was in contradiction with the expectation that the large number of final states in high-energy particle production would suppress polarization effects and showed that spin effects contribute significantly even at high-energy.
A polarizing fragmentation function, denoted by $D_{1T}^{\perp}$, has been proposed in Refs.~\cite{Boer_2008,boer2009transverse,Mulders_1996} to account for the polarized production of \Lz hyperons.
The mechanism involving the $D_{1T}^{\perp}$ function is the same as that used in the framework of the transverse-momentum-dependent unpolarized fragmentation functions (TMDs) to describe the fragmentation of an unpolarized quark into a transversely polarized hadron. 
The spin and azimuthal asymmetries observed at sufficiently large energy scales cannot be explained by asymmetries at the level of the hard partonic process, instead their origin must lie in soft processes. By maintaining an explicit dependence on the intrinsic partonic motion, TMDs account for spin and momentum correlations at the soft level, potentially explaining the observed asymmetries. Since these functions arise from soft mechanisms, they are difficult to calculate from first principles. Hence, just as with collinear parton distribution functions and fragmentation functions, one possible approach is to determine them from experimental data.
Several attempts were made to describe \Lz  polarization, both on the theoretical and experimental sides, at different accelerators and 
center-of-mass energies.  Particularly relevant are the measurements from the STAR experiment at RHIC\cite{gao2024measurement} and \belle at KEKB\cite{Guan_2019}. None of these results led to a fully satisfactory answer, and the mechanism giving rise to \Lz polarization is still unclear.
In this paper, a measurement of transverse \Lz and \Lbar polarization is presented, using the LHCb experiment in a fixed-target configuration.
The polarization is determined using proton-neon ($p$Ne) data collected in 2017 from collisions at a nucleon-nucleon center-of-mass energy of $\sqrt{s_{\rm NN}} = 68.4 \gev$, generated by a $2.5 \tev$ proton beam incident on neon nuclei at rest and corresponding to an integrated luminosity of 24.9 $\rm{nb}^{-1}$. The hyperons are reconstructed through the decays $\Lz \rightarrow p\pi^-$ and $\Lbar \rightarrow \bar{p}\pi^+$.
The results are obtained as a function of the transverse momentum (\pt) of the hyperon, the pseudorapidity ($\eta$), the rapidity ($y$) and the Feynman variable $x_\text{F} = \frac{2 \cdot p_{\mathrm{L}}}{\sqsnn} $ (where $p_{\mathrm{L}}$ is the longitudinal momentum of the particle), as a non trivial dependence of the polarization on these variables has been reported in other publications \cite{Airapetian_2014}.

\section{The LHCb detector}
The LHCb detector~\cite{LHCb-DP-2008-001,LHCb-DP-2014-002} is a single-arm forward spectrometer, designed for the study of particles containing $c$ or $b$ quarks, covering the pseudorapidity range $2 < \eta < 5$. The detector includes: a silicon-strip vertex locator (VELO), three tracking stations of silicon-strip detectors and straw drift tubes,
two ring-imaging Cherenkov detectors (RICH) that are able to discriminate between different species of charged hadrons, a calorimeter system consisting of scintillating-pad and preshower detectors, electromagnetic and hadronic calorimeters, and a muon detector composed of alternating layers of iron and multiwire proportional chambers. 
The System for Measuring the Overlap with Gas (SMOG) \cite{LHCb-PAPER-2014-047} enables the injection of gases with a pressure of $\mathcal{O}(10^{-7})$ mbar in the beam pipe section inside the VELO, allowing LHCb to operate as a fixed-target experiment. The SMOG system thus provides a unique opportunity to study proton-nucleus and nucleus-nucleus collisions with various gaseous targets using the LHC beams. In this configuration, the LHCb acceptance extends to the negative rapidity hemisphere, due to the boost induced by the high-energy proton beam, which points to  the positive-$z$ direction.

\section{Data sample and analysis strategy}
\begin{figure}[b]
  \begin{center}
    \includegraphics[width=0.49\linewidth]{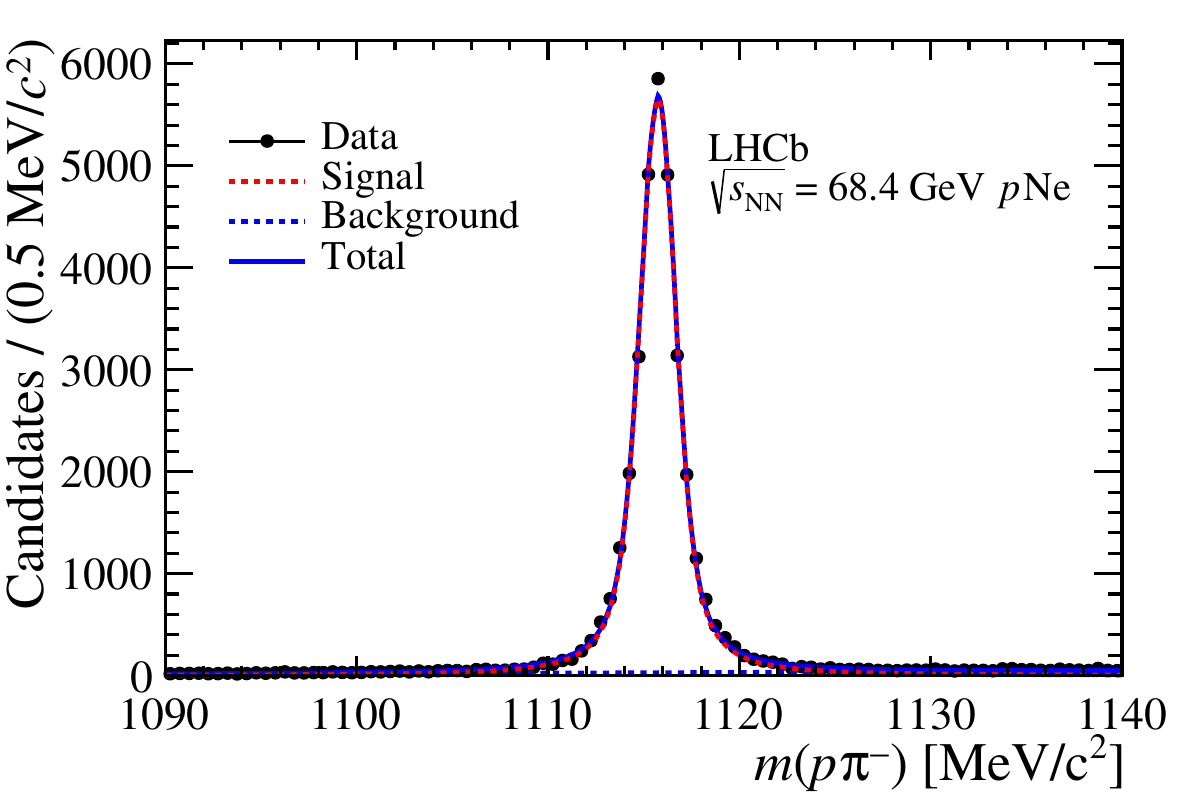}
    \includegraphics[width=0.49\linewidth]{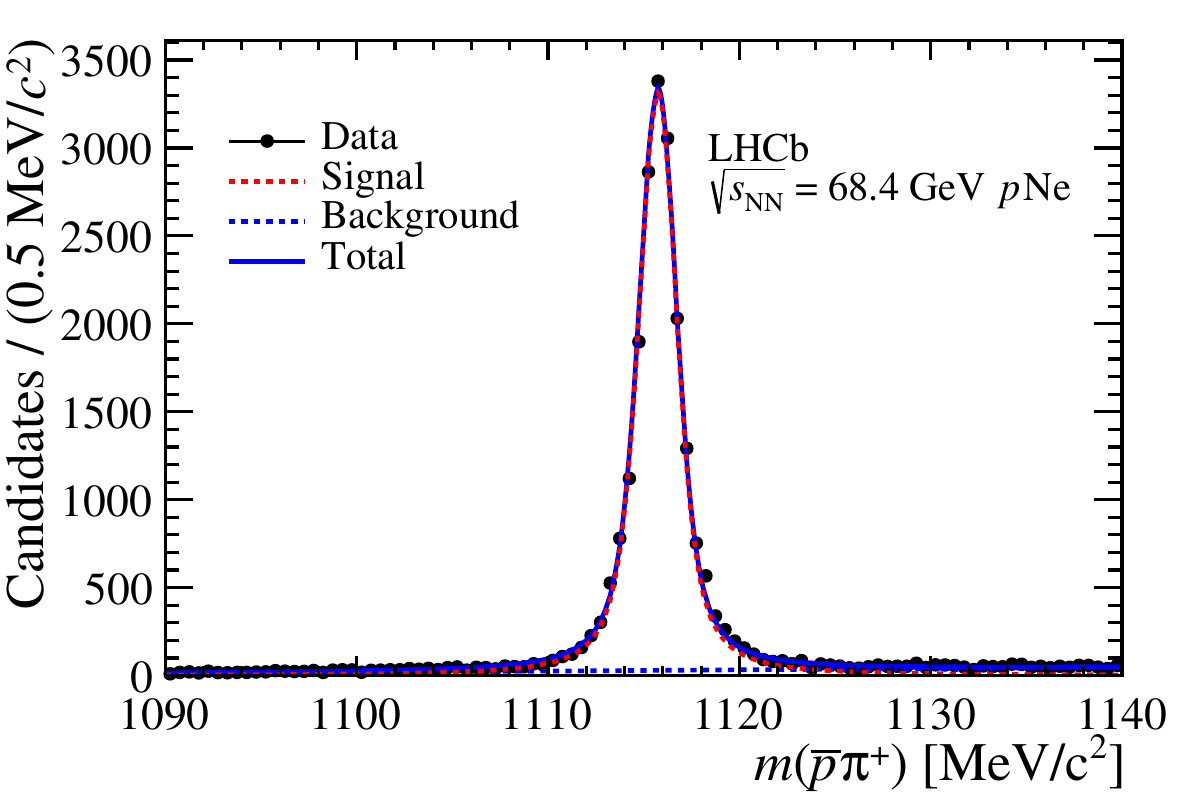}
    \vspace*{-0.5cm}
  \end{center}
  \caption{Invariant-mass distributions for (left) $\ensuremath{\PLambda}$ and (right) $\ensuremath{\offsetoverline{\PLambda}}$ candidates after all selection requirements are applied. The fit result is overlaid on the data.}
  \label{fig:MM}
\end{figure}

\begin{figure}[tb]
  \begin{center}
    \includegraphics[width=0.49\linewidth]{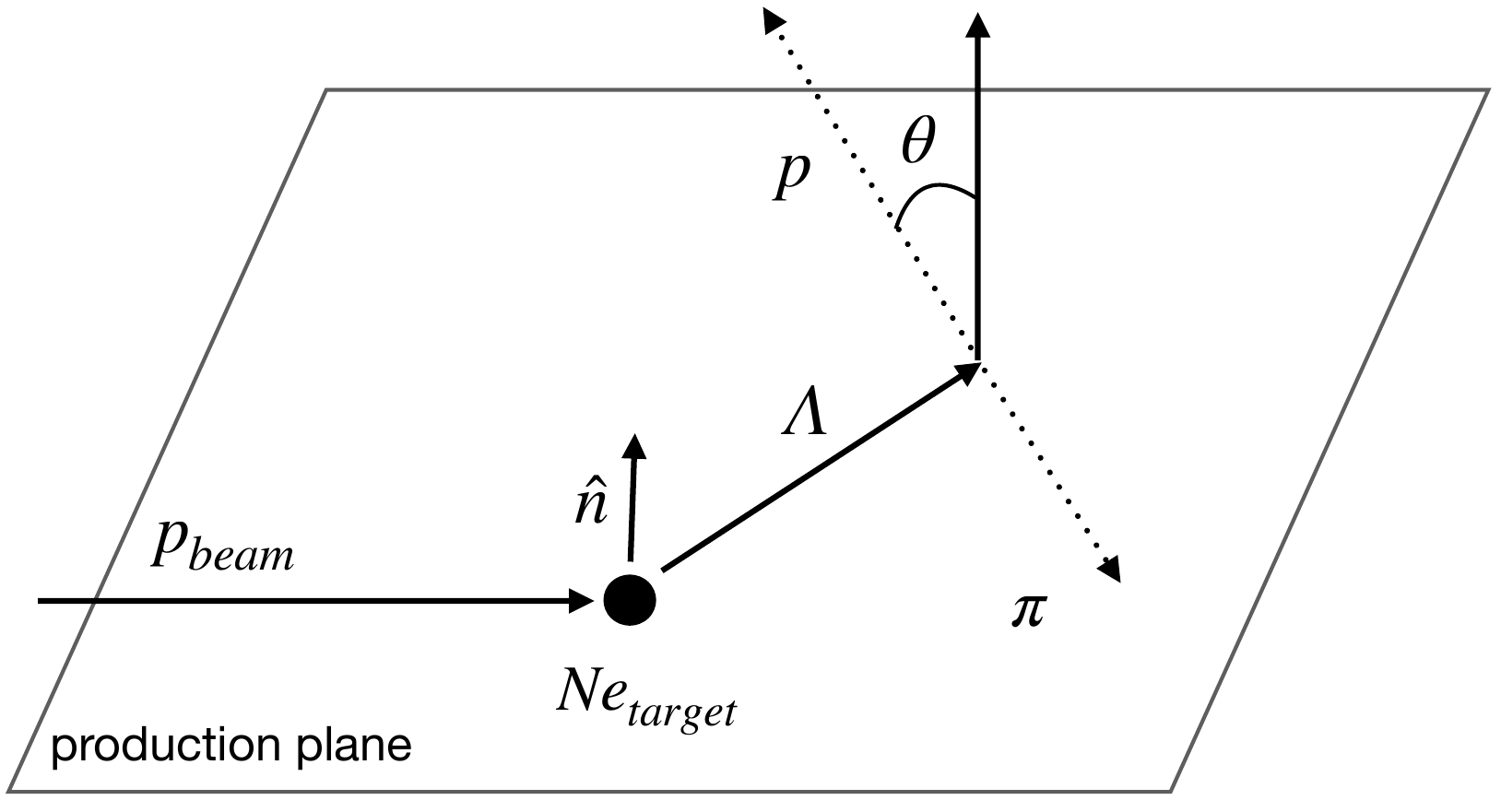}
    \vspace*{-0.5cm}
  \end{center}
  \caption{Sketch of the $\ensuremath{\PLambda}$ production and decay. The polarization vector $\hat{P_{\ensuremath{\PLambda}}}$ is aligned with the normal vector $\hat{n}$ to the $\ensuremath{\PLambda}$ production plane; $\theta$ is the angle between the momentum of the decay proton and $\hat{n}$ in the $\ensuremath{\PLambda}$ rest frame.}
  \label{fig:sketch}
\end{figure}

The $p$Ne sample was taken during the $pp$ data-taking period. Fixed-target events were collected only when a bunch in the beam pointing towards the LHCb detector crossed the interaction region without a corresponding bunch in the beam pointing to the opposite direction. To suppress the remaining $pp$ background, the $z$-coordinate of the $p$Ne primary vertex (PV) is required to lie in the fiducial region $z_{\rm{PV}} \in [-200,-100] \cup [100, 150] \, \rm{mm}$.\footnote{$z_{\rm{PV}} = 0$ is the $z$ coordinate of the center of the $pp$ interaction region.} Furthermore, events with more than four hits in the VELO stations upstream of the interaction region are rejected.
The online event selection is performed by a trigger \cite{LHCb-DP-2012-004}, that requires at least one track reconstructed in the VELO. Since the \Lz reconstruction requires two tracks, this trigger condition does not bias the measurement.
In the offline selection, the \Lz candidates are required to be reconstructed from proton and pion tracks with opposite charge, forming a vertex with a good-quality fit. 
Protons and pions are required to have a minimum transverse momentum of $100\mevc$ and a minimum momentum of $2\gevc$. Particle identification (PID) requirements based on the information from the RICH detectors are applied to select protons. Finally, to suppress the contribution of nonprompt \Lz hyperons, the impact parameter of the \Lz candidate with respect to the  PV is required to be less than $1.5\,\rm{mm}$, which reduces this contribution to about 5\%. 
Figure~\ref{fig:MM} shows the $p \pi^{-}$ and $\bar{p}\pi^{+}$ invariant-mass distributions obtained after all the selection criteria have been applied. The distributions are fitted with the convolution of a Cauchy and a Gaussian function to describe the signal shape and a first-order polynomial to model the background.
\newpage
\noindent
The decays $\Lz \rightarrow p \pi^-$ and $\Lbar \rightarrow \bar{p} \pi^+$ exhibit significant parity violation, resulting in large asymmetries in the angular distribution of their decay particles. In particular, the angular distribution of the proton in the \Lz rest frame is given by
\begin{equation}
\frac{dN}{d\cos\theta} = \frac{dN_0}{d\cos\theta}(1+\alpha P_{\Lz}\cos\theta),
\label{eq:pol}
\end{equation}
where $\theta$ is the angle between the proton momentum and the normal ($\hat{n}$) to the production plane spanned by the beam and the ${\Lz}$ momentum directions as sketched in Fig.~\ref{fig:sketch}, $\frac{dN_0}{d\cos\theta}$ is the decay distribution for unpolarized \Lz hyperons, $P_{\Lz}$ is the magnitude of the \Lz polarization, and $\alpha$ is the value of the parity-violating decay asymmetry for the channel under study of the \Lz hyperon. The magnitude of the polarization is determined from a fit to the angular distribution of the proton in 10 bins of $\cos\theta$.

\section{Simulation and efficiencies}
Efficiencies are estimated using samples of fully simulated events.
The simulated decays are reconstructed and analyzed using the same software tools as those used to process the data. In the simulation, \Lz hyperons are generated using \pythia \cite{Sjostrand:2007gs} with a specific LHCb configuration \cite{LHCb-PROC-2010-056} and with colliding-proton beam momentum equal
to the momentum per nucleon of the beam and target in the centre-of-mass frame. The decays of unstable particles are described by \evtgen \cite{Lange:2001uf}, in which final-state radiation is generated using PHOTOS \cite{Golonka_2006}.
The four-momentum of the \Lz decay particles is then embedded into $p$Ne minimum bias events that are generated with the EPOS event generator \cite{Pierog_2015}.
The interaction of the generated particles with the detector and its response are implemented using the GEANT4 toolkit \cite{Agostinelli:2002hh,Allison:2006ve} as described in Ref. \cite{LHCb-PROC-2011-006}.

After reconstruction, the simulated samples are weighted to improve agreement with data.
Weights are calculated as the ratio between the normalized data and simulated distributions as a function of $\pt$, $\eta$ and $z_{\rm{PV}}$ with
\begin{equation}
w(\pt,\, \eta, \, z_{\rm{PV}} ) = w(\pt, \, \eta) \cdot w (z_{\rm{PV}}),
\end{equation}
where $w(\pt, \eta)$ is evaluated in 6 intervals of transverse momentum between $300$ and $2500 \mevc$ and 7 intervals of pseudorapidity between 2 and 5. The weights $w(z_{\rm{PV}})$ are evaluated in 6 intervals between $-200$ and 150 $\rm{mm}$, ignoring the region between $-100$ and 100\,mm outside the fiducial region considered in the analysis.
Through the weighting procedure, the simulation is corrected on average by 6\% as a function of transverse momentum and pseudorapidity, and by 4\% as a function of $z_{\rm{PV}}$, for both \Lz and \Lbar hyperons.
Accounting for efficiency factors, Eq.~\ref{eq:pol} is modified to 
\begin{equation}
 \frac{dN}{d\cos\theta} = \frac{dN_0}{d\cos\theta}(1+\alpha P_{\Lz}\cos\theta) \times \epsilon(\cos\theta) \times \epsilon_{\rm PID},
\end{equation}
where $\epsilon_{\rm PID}$ is the particle identification efficiency for the protons, estimated from dedicated calibration data samples and computed as a function of the proton kinematics. The $\epsilon(\cos\theta)$ term is the product of acceptance, reconstruction and selection efficiencies. It is determined using simulation as $\epsilon(\cos\theta) = {f_{\rm{rec}}(\cos\theta)}/{f_{\rm{gen}}(\cos\theta)}$, where $f$ indicates the $\cos\theta$ distribution for generated candidates ($f_{\rm gen}$), without any detector effect, or for fully reconstructed candidates ($f_{\rm rec}$), with detector effects included.  
As the polarization effects are not included in the simulation, the generated $\cos\theta$ distribution is uniform. Since only the shape of the distribution is relevant, the efficiency is proportional to the reconstructed $\cos\theta$ distributions, and the generated one is ignored.
To correct the angular distributions in data, in each $\cos\theta$ bin, the normalized number of candidates in the data is divided by the normalized number of candidates in the simulated reconstructed sample.

\section{Results and systematic uncertainties}
\begin{figure}[tb]
  \begin{center}
    \includegraphics[width=0.49\linewidth]{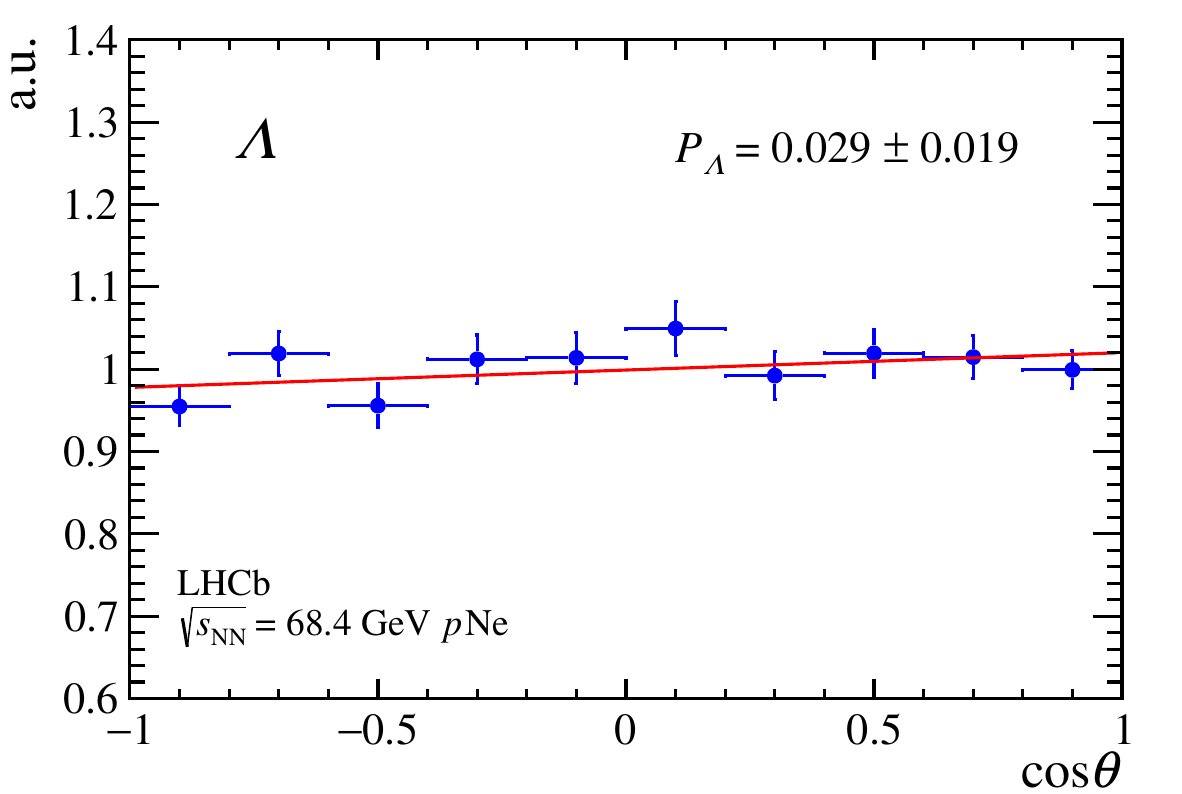}
    \includegraphics[width=0.49\linewidth]{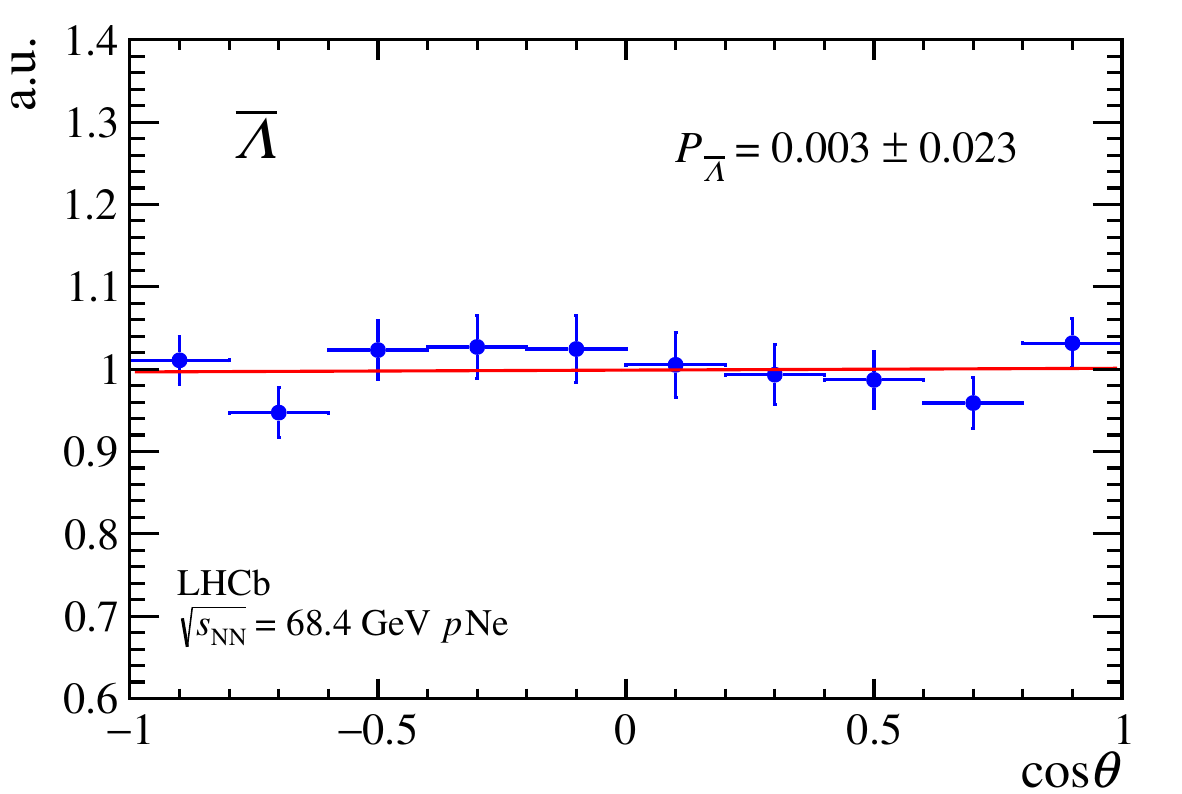}
    \vspace*{-0.5cm}
  \end{center}
  \caption{Efficiency-corrected $\cos\theta$ distributions, for (left) $\ensuremath{\PLambda}$ and (right) $\ensuremath{\offsetoverline{\PLambda}}$ hyperons. The result of the fit is overlaid (red line).}
  \label{fig:fit}
\end{figure}

The proton angular distributions, after efficiency correction, are shown in Fig.~\ref{fig:fit}. The function $f(\cos\theta) = A  (1 + \alpha P_{\Lz} \cos\theta)$ is fitted to the data distributions, where \mbox{$\alpha = 0.746 \pm 0.007$} for \Lz and \mbox{$\alpha = - 0.757 \pm 0.004$} for \Lbar are fixed to their world average value \cite{PDG2022}. The magnitude of the polarization is given by the free parameter of the fit shown in Fig.~\ref{fig:fit}.

Several sources of systematic uncertainties are considered. 
The fit to the invariant-mass distribution is repeated using a double-sided Crystal Ball function \cite{Skwarnicki:1986xj} instead of the convolution of a Cauchy and a Gaussian function for the signal shape, and a second-order polynomial instead of a first-order one for the background, and the values of the polarization are determined again. The systematic uncertainty is determined as the deviations of these results with respect to those of the default fit.
Uncertainties related to the weighting procedure applied to the simulation are taken into consideration by carrying out 100 trials, randomly varying each weight within its uncertainty, calculating new values for the polarization, and taking as systematic uncertainty the largest difference in the polarization values compared to the default one.
The choice of the variables used to weight the simulation is also considered; the uncertainty is calculated as the difference between the results obtained by using the track multiplicity and those obtained using the \Lz pseudorapidity in the calculation of the weights.
The choice of binning for the angular distributions affects the fit results. This is taken into account by repeating the polarization measurements using 5 bins instead of 10 in the angular distribution. Another contribution is associated with the estimation of PID efficiencies. An alternative approach is used, where the particle identification efficiencies are directly estimated from the simulation rather than using dedicated calibration samples.
Another systematic uncertainty contribution arises from nonprompt \Lz hyperon contamination in the data sample, which is estimated from simulation to account for 5\% of the total yield. To estimate an upper limit on the nonprompt contamination, the impact parameter requirement (which retains approximately 50\% of the nonprompt signal) is removed, the measurement is repeated and the difference with the baseline value is taken as systematic uncertainty. The systematic uncertainty due to the external parameter $\alpha$ is found to be negligible.  The total systematic uncertainty is computed as the sum in quadrature of each contribution shown in Table~\ref{tab:res_sys}. The systematic contributions are found to be small and the measurement is dominated by statistical uncertainties. The statistical effect on each systematic contribution is not negligible as reflected in the differences between \Lz and \Lbar hyperons.
The final polarization measurements are
\begin{linenomath}$$ P_{\Lz} = 0.029 \pm 0.019 \, (\rm{stat}) \pm 0.012 \, (\rm{syst}) \, ,$$
$$ P_{\Lbar} = 0.003 \pm 0.023 \, (\rm{stat}) \pm 0.014 \, (\rm{syst}) \, . $$\end{linenomath}

\begin{table}[tb]
    \centering
    \caption{Contributions of systematic uncertainties on the polarization measurement for $\ensuremath{\PLambda}$ and $\ensuremath{\offsetoverline{\PLambda}}$ hyperons.} 
    \label{tab:res_sys}
    \begin{tabular}{lcc} 
         Source&  \Lz& \Lbar\\ \hline 
         Signal estimation&  0.007& 0.001\\ 
         Background estimation&  0.001& 0.010\\
         Kinematic weights &  0.001& 0.001\\ 
         Multiplicity dependence&  0.001& 0.004\\
         Binning of $\cos\theta$ distributions&  0.007& 0.006\\ 
         PID efficiencies&  0.002& 0.005\\ 
         Nonprompt contamination&  0.005& 0.002\\ \hline
    \end{tabular}
\end{table}
\noindent

\begin{figure}[tb]
  \begin{center}
    \includegraphics[width=0.49\linewidth]{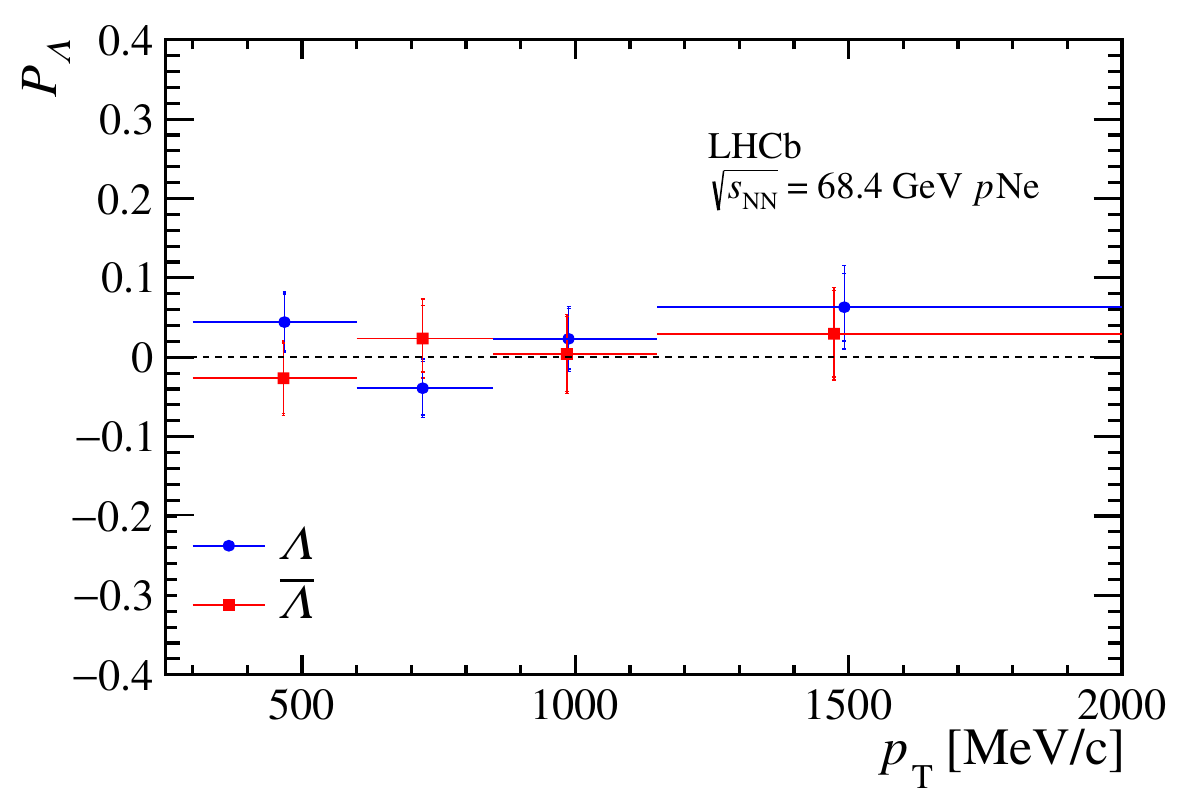}
    \includegraphics[width=0.49\linewidth]{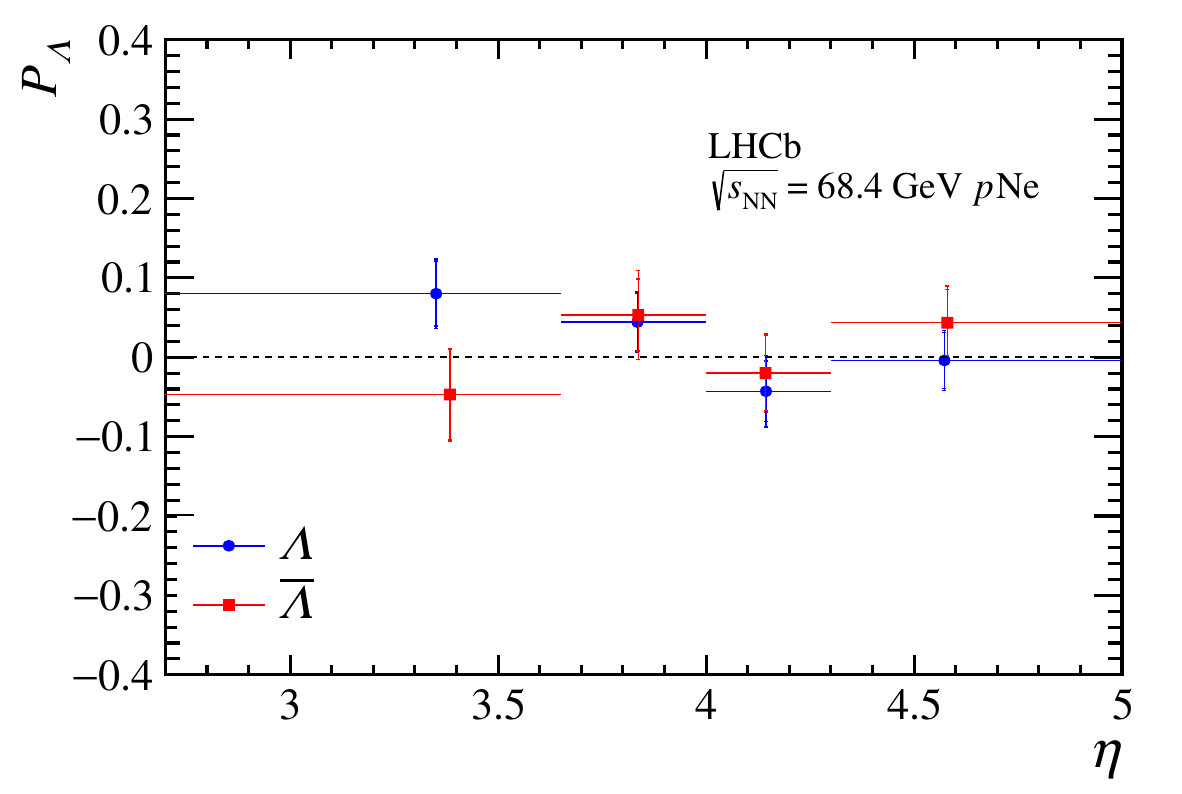}
    
    \includegraphics[width=0.49\linewidth]{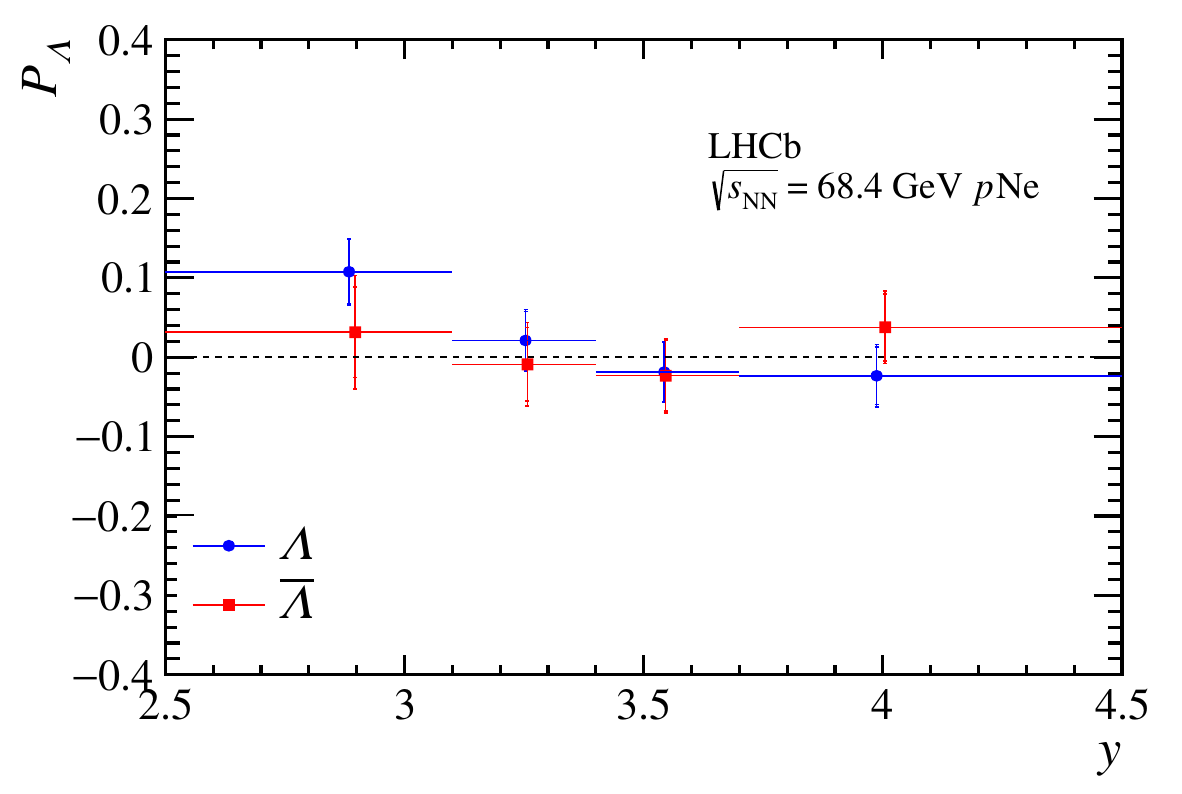}
    \includegraphics[width=0.49\linewidth]{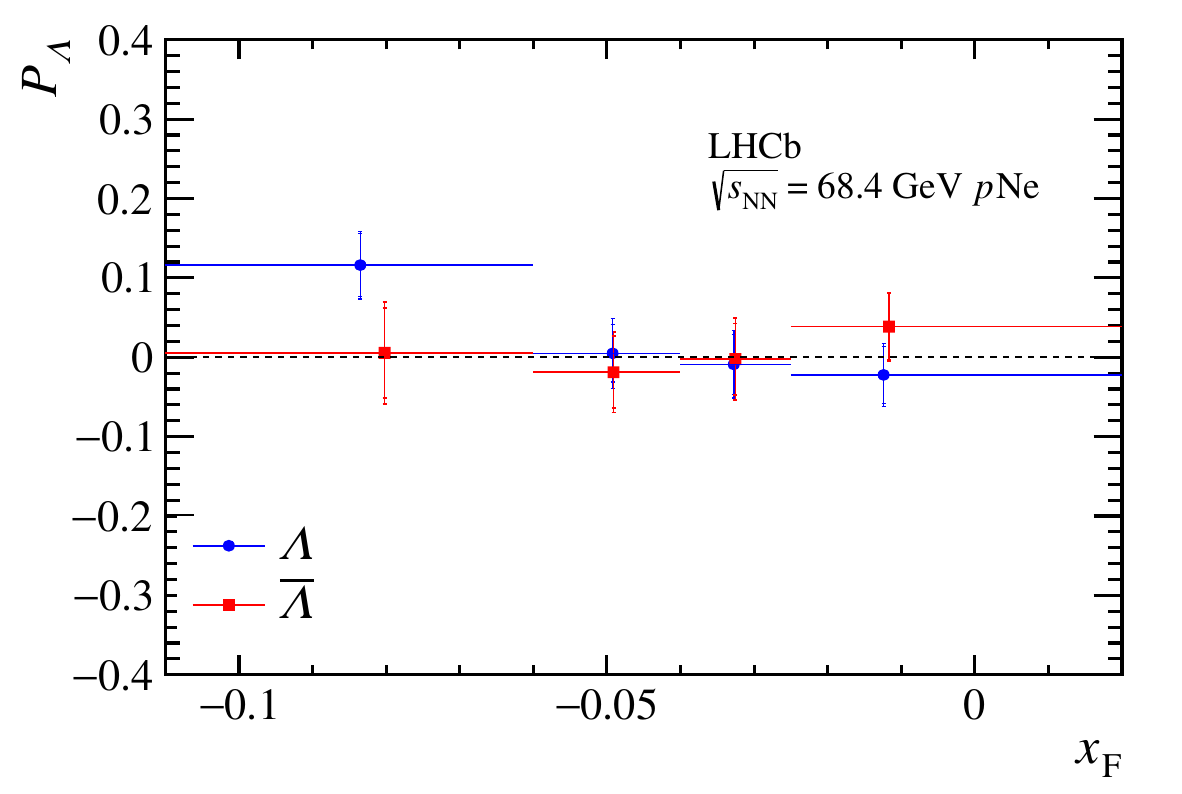}
    \vspace*{-0.5cm}
  \end{center}
  \caption{Polarization as a function of (a) $\ensuremath{p_{\mathrm{T}}}$, (b) $\eta$, (c) $x_\text{F}$ and (d) $y$. Blue (red) symbols are for $\ensuremath{\PLambda}$ ($\ensuremath{\offsetoverline{\PLambda}}$). In each plot the data is integrated over the 0.3 $<$ \ensuremath{p_{\mathrm{T}}} $<$ 3 $\ensuremath{\aunit{Ge\kern -0.1em V\!/}c}$ and/or 2 $< \eta <$ 5 kinematic range.}
  \label{fig:bins}
\end{figure}

\begin{figure}[ht]
  \begin{center}
    \includegraphics[width=0.9\linewidth]{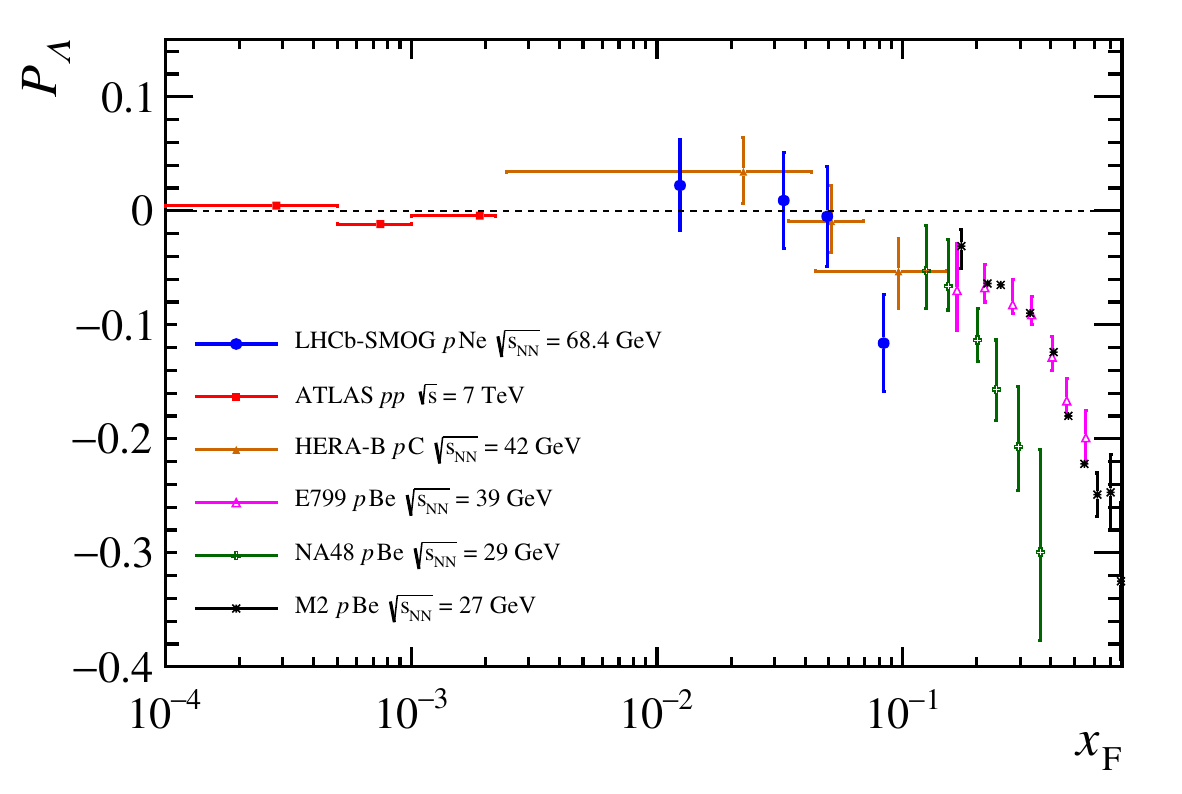}
    \vspace*{-0.5cm}
  \end{center}
  \caption{Comparison of polarization as a function of $x_\text{F}$ for $\ensuremath{\PLambda}$ hyperons obtained in experiments with different energies and with different colliding systems.}
  \label{fig:comparison}
\end{figure}

The polarization measurements have also been performed in bins of $\pt$, $\eta$, $y$ and $x_\text{F}$. The results are shown in Fig.~\ref{fig:bins} and listed in Table~\ref{tab:res_bin}.
Other experiments with different energies and collision systems have measured the \Lz polarization.
In its fixed-target configuration, the LHCb experiment covers an energy and kinematic range that is largely unexplored. Figure~\ref{fig:comparison} compares and shows the agreement between the results of this paper with measurements from other experiments, including ATLAS \cite{Aad_2015}, an experiment at the M2 beam-line at Fermilab \cite{M2}, the E799 experiment \cite{RAMBERG1994403}, NA48 \cite{na48}, and HERA-B \cite{Abt_2006}. The measurements reported here, and those from HERA-B, cover negative values of $x_\text{F}$, so the results are first transformed using the following symmetry of the transverse polarization $P_{\Lz}(-x_{\rm F}) = - P_{\Lz}(x_{\rm F})$, and then compared with the other measurements.

\begin{table}[ht]
	\centering
	\caption{\label{tab:res_bin}Polarization in bins of $\ensuremath{p_{\mathrm{T}}}$, $\eta$, $x_\text{F}$, and $y$ for $\ensuremath{\PLambda}$ and $\ensuremath{\offsetoverline{\PLambda}}$. The first uncertainties are statistical and the second are systematic.}
	\small
    \begin{tabular}{lcc}
        &$P_{\Lz}$&$P_{\Lbar}$\\
        \hline
        
		  \pt $[\mevc ]$ & & \\

		$[300,600]  $& \phantom{0} $0.044 \pm 0.035  \pm 0.013 $     & $-0.026\pm 0.044\pm 0.015$\\
        $[600,850]  $& $- 0.039 \pm 0.033 \pm 0.016$     & \phantom{0} $0.023 \pm 0.042  \pm 0.027$\\
		$[850,1150] $& \phantom{0} $0.023 \pm 0.038 \pm 0.015$     & \phantom{0} $0.004 \pm 0.047 \pm 0.015$\\
        $[1150,3000]$& \phantom{0} $0.063 \pm 0.042 \pm 0.031$     & \phantom{0} $0.029 \pm 0.054 \pm 0.021$\\
		\hline 
        
		$\eta$ &  &\\
        
		$[2.00,3.65]$ & \phantom{0} $0.080 \pm 0.040 \pm 0.016$ & $- 0.047 \pm 0.057 \pm 0.011$\\
		$[3.65,4.00]$ & \phantom{0} $0.044 \pm 0.036 \pm 0.012$ & \phantom{0} $0.053 \pm 0.045 \pm 0.033$\\
		$[4.00,4.30] $ & $- 0.043 \pm 0.038 \pm 0.024$ & $- 0.020 \pm 0.047 \pm 0.014$\\
		$[4.30,5.00] $ & $- 0.004 \pm 0.035 \pm 0.014$ & \phantom{0} $0.043 \pm 0.042 \pm 0.020$\\
		\hline
        
		$x_\text{F}$ &   &\\
        
		$[-0.250,-0.060]$ & \phantom{0} $0.116 \pm 0.040 \pm 0.015$ & \phantom{0} $0.006 \pm 0.057 \pm 0.030$\\ 
		$[-0.060,-0.040]$ & \phantom{0} $0.005 \pm 0.036 \pm 0.025$ & $- 0.018 \pm 0.045 \pm 0.023$\\
		$[-0.040,-0.025]$ & $- 0.009 \pm 0.037 \pm 0.020 $ & \phantom{0} $0.002 \pm 0.045 \pm 0.026$\\
		$[-0.025,0.100] $ & $- 0.022 \pm 0.036 \pm 0.017$ & \phantom{0} $0.038 \pm 0.042 \pm 0.008$\\
	    \hline
        
		$y$ &   &\\
        
		  $[2.0,3.1]$ & \phantom{0} $0.107  \pm 0.040 \pm 0.011$     & \phantom{0} $0.031 \pm 0.057 \pm 0.044$\\ 
		$[3.1,3.4]$ & \phantom{0} $0.021  \pm 0.036 \pm 0.013$     & $- 0.009 \pm 0.046 \pm 0.025$\\
		$[3.4,3.7]$ & $- 0.019 \pm 0.037 \pm 0.010$     & $- 0.023 \pm 0.045 \pm 0.013$\\
		$[3.7,5.0]$ & $- 0.023 \pm 0.036 \pm 0.016$     & \phantom{0} $0.038  \pm 0.042 \pm 0.018$\\
 \end{tabular}
\end{table}

\section{Conclusions}
A measurement of transverse polarization of \Lz and \Lbar hyperons in $p$Ne collisions at $\sqsnn=68.4 \gev$ by the LHCb experiment is presented. This analysis exploits the innovative and unique fixed-target apparatus at LHC. The measurement in a new collision system and region of phase space can provide additional insights on the mechanism of \Lz polarization, which can still not be calculated in quantum chromodynamics.
The polarization is measured through the decay $\Lz \to p \pi^-$ and its charge conjugate, and is studied both as integrated values and in different bins of four kinematic variables: $\pt$, $\eta$, $x_\text{F}$, and $y$. 

The integrated results are \begin{linenomath}$$ P_{\Lz} = 0.029 \pm 0.019 \, (\rm{stat}) \pm 0.012 \, (\rm{syst}) \, , $$ $$ P_{\Lbar} = 0.003 \pm 0.023 \, (\rm{stat}) \pm 0.014 \,(\rm{syst}) \,. $$\end{linenomath}
The polarization values obtained in this analysis are compatible with previous measurements, in particular with the HERA-B results which cover a similar $x_\text{F}$ interval. The agreement is noteworthy considering the different experiments and colliding systems.

The LHCb fixed-target program has become particularly active since the installation of the SMOG2 system, which increases data luminosity by more than two orders of magnitude compared to the SMOG system used in this analysis. With the new data, we expect to explore different target species and improve our reach into the lower negative regions of $x_\text{F}$ which have been poorly explored so far.

%% file: acknowledgements.tex
\section*{Acknowledgements}
%
% These Acknowledgements valid from 3-May-2019
%

We express our gratitude to our colleagues in the CERN
accelerator departments for the excellent performance of the LHC. We
thank the technical and administrative staff at the LHCb
institutes.
We acknowledge support from CERN and from the national agencies:
CAPES, CNPq, FAPERJ and FINEP (Brazil); 
MOST and NSFC (China); 
CNRS/IN2P3 (France); 
BMBF, DFG and MPG (Germany); 
INFN (Italy); 
NWO (Netherlands); 
MNiSW and NCN (Poland); 
MCID/IFA (Romania); 
%MSHE (Russia); 
MICINN (Spain); 
SNSF and SER (Switzerland); 
NASU (Ukraine); 
STFC (United Kingdom); 
DOE NP and NSF (USA).
We acknowledge the computing resources that are provided by CERN, IN2P3
(France), KIT and DESY (Germany), INFN (Italy), SURF (Netherlands),
PIC (Spain), GridPP (United Kingdom), 
%RRCKI and Yandex LLC (Russia), 
CSCS (Switzerland), IFIN-HH (Romania), CBPF (Brazil),
and Polish WLCG (Poland).
We are indebted to the communities behind the multiple open-source
software packages on which we depend.
Individual groups or members have received support from
ARC and ARDC (Australia);
Key Research Program of Frontier Sciences of CAS, CAS PIFI, CAS CCEPP, 
Fundamental Research Funds for the Central Universities, 
and Sci. \& Tech. Program of Guangzhou (China);
Minciencias (Colombia);
EPLANET, Marie Sk\l{}odowska-Curie Actions, ERC and NextGenerationEU (European Union);
A*MIDEX, ANR, IPhU and Labex P2IO, and R\'{e}gion Auvergne-Rh\^{o}ne-Alpes (France);
%RFBR, RSF and Yandex LLC (Russia);
AvH Foundation (Germany);
ICSC (Italy); 
GVA, XuntaGal, GENCAT, Inditex, InTalent and Prog.~Atracci\'on Talento, CM (Spain);
SRC (Sweden);
the Leverhulme Trust, the Royal Society
 and UKRI (United Kingdom).

%% file: Authorship_LHCb-PAPER-2024-009.tex
% LHCb collaboration author list
% Data extracted on May 13th, 2024 at 3:59pm for paper reference LHCb-PAPER-2024-009
\centerline
{\large\bf LHCb collaboration}
\begin
{flushleft}
\small
R.~Aaij$^{36}$\lhcborcid{0000-0003-0533-1952},
A.S.W.~Abdelmotteleb$^{55}$\lhcborcid{0000-0001-7905-0542},
C.~Abellan~Beteta$^{49}$,
F.~Abudin{\'e}n$^{55}$\lhcborcid{0000-0002-6737-3528},
T.~Ackernley$^{59}$\lhcborcid{0000-0002-5951-3498},
A. A. ~Adefisoye$^{67}$\lhcborcid{0000-0003-2448-1550},
B.~Adeva$^{45}$\lhcborcid{0000-0001-9756-3712},
M.~Adinolfi$^{53}$\lhcborcid{0000-0002-1326-1264},
P.~Adlarson$^{79}$\lhcborcid{0000-0001-6280-3851},
C.~Agapopoulou$^{13}$\lhcborcid{0000-0002-2368-0147}
% ...

{\footnotesize \it

% ...
$^{13}$Universit{\'e} Paris-Saclay, CNRS/IN2P3, IJCLab, Orsay, France\\
% ...
$^{36}$Nikhef National Institute for Subatomic Physics, Amsterdam, Netherlands\\
% ...
$^{45}$Instituto Galego de F{\'\i}sica de Altas Enerx{\'\i}as (IGFAE), Universidade de Santiago de Compostela, Santiago de Compostela, Spain\\
% ...
$^{49}$Physik-Institut, Universit{\"a}t Z{\"u}rich, Z{\"u}rich, Switzerland\\
% ...
$^{53}$H.H. Wills Physics Laboratory, University of Bristol, Bristol, United Kingdom\\
% ...
$^{55}$Department of Physics, University of Warwick, Coventry, United Kingdom\\
% ...
$^{58}$School of Physics and Astronomy, University of Glasgow, Glasgow, United Kingdom\\
$^{59}$Oliver Lodge Laboratory, University of Liverpool, Liverpool, United Kingdom\\
% ...
$^{67}$Syracuse University, Syracuse, NY, United States\\
% ...
$^{79}$Department of Physics and Astronomy, Uppsala University, Uppsala, Sweden, associated to $^{58}$\\
% ...
}
\end{flushleft}